\documentclass[reqno,12pt]{article}
\usepackage{amsmath,amsfonts,amssymb,amsthm,amstext,amscd,eucal,xcolor}
\usepackage[all]{xy}
\usepackage{amsmath, amssymb}
\usepackage{mathtools}
\newcommand{\mathsym}[1]{{}} 
\usepackage[colorlinks=true,
            linkcolor=blue,
            urlcolor=blue,
            citecolor=blue]{hyperref}
\usepackage{enumerate}
\usepackage{slashed}

\usepackage{listings}
\DeclareMathAlphabet{\pazocal}{OMS}{zplm}{m}{n}
\usepackage{graphicx}
\usepackage{amsmath} \usepackage{amssymb}
\usepackage{bm}
\usepackage{cite,hyperref}
\usepackage{mathtools}
\usepackage{amsmath} \usepackage{amssymb}
\usepackage{bm}
\usepackage{graphicx}
\usepackage[active]{srcltx}
 \usepackage{color}
\definecolor{DeepPink}{rgb}{1.00,0.08,0.58}
\definecolor{DeepPink2}{rgb}{0.93,0.07,0.54}
\makeatletter \@addtoreset{equation}{section}

\makeatletter\renewcommand\section{\@startsection {section}{1}{\z@}%
                                   {-3.5ex \@plus -1ex \@minus -.2ex}
                                   {2.3ex \@plus.2ex}%
                                   {\normalfont\large\bfseries}}
\renewcommand\subsection{\@startsection{subsection}{2}{\z@}%
                                     {-3.25ex\@plus -1ex \@minus -.2ex}%
                                     {1.5ex \@plus .2ex}%
                                     {\normalfont\bfseries}}

\DeclareMathAlphabet{\mathcal}{OMS}{cmsy}{b}{n}

\makeatother

\newcommand{\email}[1]{\footnote{E-mail: \href{mailto:#1}{#1}}}

\begin{document}

\title{\bf\Large{  Induced non-Abelian Chern-Simons effective action in\\ very special relativity }}

\author{\textbf{Z. Haghgouyan\email{z$_{_{-}}$haghgooyan@sbu.ac.ir} $^{a}$, M.~Ghasemkhani\email{m$_{_{-}}$ghasemkhani@sbu.ac.ir } $^{a}$, R.~Bufalo\email{rodrigo.bufalo@ufla.br} $^{b}$, and A.~Soto\email{arsoto1@uc.cl} $^{c}$ }\\\\
\textit{\small $^{a}$  Department of Physics, Shahid Beheshti University, 1983969411, Tehran, Iran}\\
\textit{$^{b}$ \small Departamento de F\'isica, Universidade Federal de Lavras,}\\
\textit{ \small Caixa Postal 3037, 37200-900 Lavras, MG, Brazil}\\
\textit{\small$^c$ School of Mathematics, Statistics and Physics, Newcastle University,}\\
 \textit{ \small Newcastle upon Tyne, NE1 7RU, UK}\\
}

\maketitle
\date{}

\begin{abstract}
In this paper, we study the one-loop induced effective action for a non-abelian gauge field in the very special relativity (VSR) framework in a $(2+1)$-dimensional spacetime.
We show that there are new graphs contributing to the amplitudes of gluon $n$-point functions, e.g. $\langle GG\rangle$, $\langle GGG \rangle$ and
$\langle GGGG \rangle$, originating from the non-local couplings generated in VSR.
Using the one-loop analysis of the relevant graphs, we evaluate the VSR corrections to the induced kinetic term of the non-abelian gauge field action in three dimensions.
Next, by applying the Ward identity, we determine the general tensorial structure for the amplitude of $\langle GG\rangle$ in VSR, which is valid at any order of the loop analysis.
Moreover, we discuss the leading VSR corrections to the non-abelian Chern-Simons and Yang-Mills action through the analysis of $\langle GG\rangle$, $\langle GGG \rangle$ and $\langle GGGG \rangle$.
In order to highlight the VSR effects, we present the VSR modification of non-abelian Chern-Simons effective action and then obtain
 the equation of motion for the non-abelian gauge field. As a result, we observe that the well-known pure gauge solution does not hold in the presence of VSR effects.
\end{abstract}


\setcounter{footnote}{0}
\renewcommand{\baselinestretch}{1.05}  

\newpage

\section{Introduction}

Although the Standard Model is an important theoretical framework for learning about the particles and their interactions, it is not a perfect or a complete theory.
For example, the recent experiment on the muon's anomalous magnetic moment has revealed the necessity of exploring new physics \cite{Muong-2:2021ojo}.
However, due to the success of the Standard Model in the prediction of the Higgs boson and the good accuracy of the electron magnetic moment, it is important to remark that any other chosen framework in certain limit should contain the Standard Model \cite{Beacham:2019nyx}.

On the other hand, up to this date, the particle content of the universe seems to be the same predicted by the Standard Model.
Under this consideration, the very special relativity (VSR) is a framework which has got attention due to the possibility of getting neutrino mass with the same number of particles present in the Standard Model \cite{Cohen:2006ky,Cohen:2006ir}.
The main statement of this proposal is that the nature is invariant under a subgroup of the Lorentz group, which preserves the basic elements of the special relativity. A rich phenomenology and theoretical considerations have been studied in several works \cite{Dunn:2006xk,Alfaro:2015fha,Lee:2015tcc,Nayak:2016zed,Alfaro:2017umk,Alfaro:2019koq,Bufalo:2019qot,Cheon:2009zx,Alfaro:2013uva,Bufalo:2016lfq,Alfaro:2019snr,Bufalo:2020cst}.

The main group used in the formulation of VSR models in the four-dimensional case is the SIM(2), which is a four-parameter group without any invariant tensors \cite{Cohen:2006ky,Cohen:2006ir}.
However, this group preserves the direction of a null vector $n$, which transforms under a general SIM(2) transformation as $n\rightarrow e^\phi n$, where $\phi$ is a phase.
 This vector defines a privileged direction and allows to construct the Lorentz violating terms containing the ratios of contractions of this null vector with kinematical variables.

Although the construction of any model in (realistic) four dimensions is very important, we observe
 major interest in lower-dimensional models existing in the literature since the 1980's \cite{Deser:1981wh,Witten:1988hc}.
These planar field theory models in $(2+1)$-dimensions have attracted many attentions in condensed matter physics such as graphene and the topological models considerations as in Chern-Simons models (see, for example, \cite{Marino:2017ckg} and references therein).

Until this moment, there are not many studies of VSR in lower dimensions.
In ref.~\cite{Alfaro:2019snr}, it was shown that we can construct the VSR-like terms in $(1+1)$-dimensions.
Also, in $(2+1)$-dimensions, it is possible to have a VSR construction with SIM(1) group (see \cite{Vohanka:2014lsa} for more details).
In the three dimensional version, the Maxwell-Chern Simons electrodynamics was already studied \cite{Bufalo:2016lfq,Bufalo:2020cst}.

In ref.~\cite{Bufalo:2020cst}, we have examined the VSR contributions to the induced Maxwell-Chern-Simons effective action.
The construction of the effective action in the abelian case \cite{Bufalo:2020cst} allowed us to investigate the contribution of the VSR terms to the Hall's conductivity.
Another important result of \cite{Bufalo:2020cst} is that Furry's theorem also holds in the VSR setup.
Therefore, the three-point function (corresponding to triangle graphs) at one-loop order does not contribute to the abelian gauge field effective action, as a result of the charge-conjugation invariance.

On the other hand, it is known that in a $(2+1)$-dimensional non-abelian gauge theory, the triangle diagrams contribute to the Chern-Simons self-coupling terms, a completely different situation when compared with the Abelian case (for a pedagogical review on these topics, see \cite{Dunne:1998qy}).

The VSR non-abelian setup can be useful to examine the validity of Furry's theorem and then compare the result with the abelian version \cite{Bufalo:2020cst}.
Moreover, since our main interest will be focused on the VSR modifications to the pure non-abelian Chern-Simons theory, we will evaluate the VSR corrections to the gauge field equation of motion and show that the (well-known) pure gauge solution does not hold for this case. The main goal of this work is to find the gluon one-loop effective action by integrating out the fermionic fields in the presence of VSR effects. Hence, we will construct the VSR generalization of a three-dimensional model including the interaction of fermionic matter with an external non-abelian gauge field.

Therefore, the outline of the paper is as follows.
In section \ref{sec2}, we will introduce the main aspects of the VSR gauge invariance. Then, we propose a Lagrangian density for the fermionic matter interacting with an external non-abelian gauge field $G_\mu ^a$, and also present the relevant Feynman rules for the case of study.
We start our main analysis in section \ref{sec3}, where we compute the two-point function $\langle G G \rangle$ and examine its general tensorial form.
In section \ref{sec4}, we compute the three-point function and obtain the leading VSR contribution to the effective action at order $(m^2/m_q^2)$ (corrections to the Chern-Simons self-coupling).
In section \ref{sec5}, we evaluate the low-energy limit of the amplitude $\langle GGGG \rangle$ in the non-VSR regime, which corresponds to the standard four-gluon vertex in ordinary quantum chromodynamics (QCD). Also, we discuss the graphs  contributing to the VSR modifications of the 4-gluon self-coupling in terms of $\langle GGGG \rangle$ qualitatively.
In addition, we consider the VSR Chern-Simons effective action at order $(m^2/m_q^2)$ to compute the equation of motion for the gauge field $G_\mu ^a$.
Finally, we will present our conclusion and final remarks in section \ref{conc}.

\section{The model}
\label{sec2}
This section will present the basic ingredients for our analysis in the VSR setup.
In the first part, we will review some aspects of the non-Abelian gauge field under this formalism. A detailed description of this topic can be found in references \cite{Alfaro:2013uva,Alfaro:2015fha} that we will follow closely.

In the VSR formalism, we can write the free fermionic Lagrangian as \cite{Cohen:2006ir}
\begin{equation}
\pazocal{L}_{_{0}} = \bar\psi\Big(i\slashed{\partial}+\frac{im^{2}}{2}\frac{\slashed{n}}{n.\partial}-m_{q}\Big)\psi ,
\end{equation}
where $m_q$ is the standard mass term for the fermion (quark).
Notice that the non-local term $\frac{\slashed{n}}{n.\partial}$ is not Lorentz invariant, but it is invariant under SIM(2) group in four dimensions or SIM(1) in three dimensions.
In three dimensions, the null vector $n$ is defined as $n_{\mu}=\left(1,0,1\right)$.
Also, the quantity $m$ is a new VSR parameter with units of mass that will measure the deviation from the Standard Model. We observe that in the case of $m\to 0$ we obtain the standard Dirac Lagrangian for a fermion.

We consider that the fermion possesses a global symmetry given by a non-abelian group.
Thus, its transformation is
\begin{equation}
\delta\psi=i\chi\psi.
\end{equation}
Now, if we promote this symmetry to a local one then the covariant derivative operator $D_{\mu}$ is defined by \cite{Alfaro:2013uva} ( for abelian case see \cite{Cheon:2009zx})
\begin{equation} \label{eq100}
D_\mu\psi=\partial_\mu \psi-i g G_\mu\psi+i g\frac{m^2}{2}\bigg(\frac{1}{(n.\partial)^2}n.G\bigg)\psi,
\end{equation}
which satisfies the following transformation law for the charged fields
\begin{equation}
\delta(D_\mu\psi)=i\chi D_\mu\psi.
\end{equation}
It defines the gauge transformation law for the associated non-abelian field $G$ \cite{Alfaro:2013uva,Alfaro:2015fha} as below
\begin{equation}
\delta G_\mu=\partial_\mu\chi-i g[G_\mu,\chi]-i g\frac{m^2}{2}n_\mu\Big[\chi,\frac{1}{(n.\partial)^2}
(n.G)\Big]+\frac{m^2}{2}\frac{n_\mu}{(n.\partial)}\chi-i g \frac{m^2}{2}\frac{n_\mu}{(n.\partial)^2}n.[G,\chi]. \label{eq105}
\end{equation}
According to the ref. \cite{Alfaro:2013uva}, we can introduce a wiggle covariant derivative
\begin{equation} \label{eq110}
\tilde{D}_\mu\psi=D_\mu\psi+\frac{m^2}{2}\frac{n_\mu}{n.D}\psi ,
\end{equation}
which reproduces, in the noninteracting case $g \to 0$, the wiggle derivative $\tilde{\partial}_\mu =\partial_\mu +\frac{m^2}{2}\frac{n_\mu}{n.\partial}$.

Furthermore, in \cite{Alfaro:2015fha}, it was shown that a field redefinition $G_\mu\to G_\mu-\frac{m^2}{2}\frac{n_\mu}{(n.\partial)^2}(n.G)$ eliminates the term proportional to $m^{2}$ in Eq.~\eqref{eq100}, without any changes in physical observable.
This observation implies that, instead of considering \eqref{eq100}, we can work with ordinary covariant derivative
\begin{equation}
\label{covariantd}
D_\mu=\partial_\mu-i g G_\mu,
\end{equation}
as a minimal coupling between the gauge and charged fields, but with interactions defined in terms of the VSR invariant wiggle operator \eqref{eq110}. Hence, using the above argument, we can construct the VSR and gauge invariant Lagrangian for the fermion fields as
\begin{equation}
\pazocal{L} = \bar\psi\Big(i\slashed{D}+\frac{im^{2}}{2}\frac{\slashed{n}}{n.D}-m_{q}\Big)\psi.
\label{eq:1}
\end{equation}
Again, we notice that in the non-VSR limit, $m\rightarrow 0$, we recover the standard Lagrangian for the fermionic fields in the presence of an external gauge field $G$.

In order to find the general structure of the gluon's one-loop effective action, we use the path-integral formalism by integrating out the fermionic fields as
\begin{equation}
e^{i\Gamma[G]}=\int D\bar\psi D\psi \,e^{i\pazocal{S}_{F}},
\end{equation}
where $\pazocal{S}_{F}$ is the fermionic action, defined as $\pazocal{S}_{F}=\int d^{3}x\,\pazocal{L}$. Using the relevant techniques in path-integral formalism, the gluon's one-loop effective action $\Gamma[G]$ is given by
\begin{equation}
\Gamma[G]=-i{\rm Tr}\ln\Big(i\slashed{D}+\frac{im^{2}}{2}\frac{\slashed{n}}{n.D}-m_{q}\Big).
\label{eq:1-2}
\end{equation}
We observe that the term $1/(n.D)$ in \eqref{eq:1} is non-local and its expansion introduces an infinite number of interactions in the coupling $g$.
Due to our interest in the perturbative computation, the effective action \eqref{eq:1-2} can be equivalently written as
\begin{equation}
\Gamma[G]=\sum_{n=1}^{\infty}\int d^{3}x_{1}\ldots \int d^{3}x_{n}~\Gamma^{a_{1}\ldots a_{n}}_{\mu_{1}\ldots \mu_{n}}
(x_{1},\ldots, x_{n})~G_{\mu_{1}}^{a_{1}}(x_{1})\ldots G_{\mu_{n}}^{a_{n}}(x_{n}).
\label{eq:1-3}
\end{equation}
Here, $\Gamma^{a_{1}\ldots a_{n}}_{\mu_{1}\ldots \mu_{n}}$ is the gluon's $n$-point function.
To determine the explicit form of this function, we can compare the series form of \eqref{eq:1-2} with \eqref{eq:1-3}, obtaining thus
\begin{equation}
\Gamma^{a_{1}\ldots a_{n}}_{\mu_{1}\ldots \mu_{n}}(x_{1},\ldots, x_{n})=-\frac{g^{n}}{n}\int\prod_{i=1}^{n}\frac{d^{3}p_{i}}{(2\pi)^{3}}
~\delta(\sum\limits_{i=1}^{n} p_{i})~e^{i\sum\limits_{i=1}^{n}p_{i}.x_{i}}~
\Xi^{a_{1}\ldots a_{n}}_{\mu_{1}\ldots \mu_{n}}(p_{1},\ldots,p_{n}),
\label{eq:1-4}
\end{equation}
where  $\Xi^{a_{1}\ldots a_{n}}_{\mu_{1}\ldots \mu_{n}}$ represents the Feynman amplitude with $n$ gluon's external legs.

\begin{figure}[t]
\vspace{-1.2cm}
\includegraphics[height=5\baselineskip]{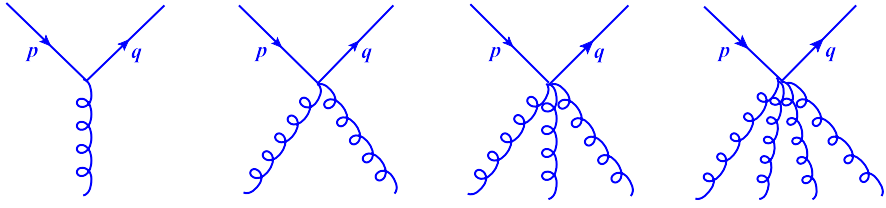}
 \centering\caption{Feynman graphs of the quark sector up to order $g^{4}$.}
\label{fig:vertices}
\end{figure}

Due to our purposes, we will only consider the vertices containing up to four gluon legs in \eqref{eq:1}.
Hence, expanding the non-local term in the Lagrangian density \eqref{eq:1}, we get
\begin{align}
\label{lagrangianexp}
\pazocal{L}=& \bar\psi\Big[i\slashed{\partial}+\frac{im^{2}}{2}\frac{\slashed{n}}{n.\partial}-m_{q}\Big]\psi
+g\bar\psi\Big[\slashed{G}-\frac{m^{2}}{2}\frac{\displaystyle{\slashed{n}}}{n.\partial}(n.G)\frac{1}{n.\partial}\Big]\psi \cr
&
-ig^{2}\bar\psi\Big[\frac{m^{2}}{2}\frac{\slashed{n}}{n.\partial}(n.G)\frac{1}{n.\partial}
(n.G)\frac{1}{n.\partial}\Big]\psi\cr
&+g^{3}\bar\psi\Big[\frac{m^{2}}{2}\frac{\slashed{n}}{n.\partial}(n.G)\frac{1}{n.\partial}
(n.G)\frac{1}{n.\partial}(n.G)\frac{1}{n.\partial}\Big]\psi+\cdots
\end{align}
The first term corresponds to the free quark Lagrangian in the VSR framework.
Thus, the free quark propagator is readily found as
\begin{equation}
S^{ij}(p)=\frac{i\left(\displaystyle{\not}\widetilde{p}+m_{q}\right)}{\widetilde{p}^{2}-m_{q}^{2}}\delta^{ij},
\end{equation}
where $\widetilde{p}_{\mu}=p_{\mu}-\frac{1}{2}\frac{m^{2}}{n.p}n_{\mu}$ and $(i,j)$ refer to the fundamental indices of the $SU(3)$ group.
We define the fermionic dispersion relation $\widetilde{p}^{2}=m_{q}^{2}$, or equivalently as $p^{2}=\mu^{2}$, where $\mu^2 = m^{2}_{q} +m^2$ is the modified quark mass.
The second term of \eqref{lagrangianexp} corresponds to the standard vertex $\langle\bar{\psi}\psi G\rangle$ added by a VSR contribution, and the next terms  produce the vertices with $n>1$ gluon legs, which are purely VSR effects.
The relevant Feynman rules can be found from the path integral formalism, where we write the spinor indices and the gauge field in terms of the $SU(3)$ gauge group generators as $G_\mu=\sum\limits_{a=1}^{8}G^a_\mu T^a$.
Hence, the interaction part of the Lagrangian density \eqref{lagrangianexp} reads
\begin{align}
{\pazocal{L}}_{int}
=&
g\bar\psi_{i}\Big[\displaystyle{\not}G^{a}-\frac{m^{2}}{2}\frac{\displaystyle{\not}n}{n.\partial}(n.G^{a})
\frac{1}{n.\partial}\Big](T^{a})_{ij}
\psi_{j} \cr
&-ig^{2}\bar\psi_{i}\Big[\frac{m^{2}}{2}\frac{\slashed{n}}{n.\partial}(n.G^{a})\frac{1}{n.\partial}
(n.G^{b})\frac{1}{n.\partial}\Big](T^{a}T^{b})_{ij}\psi_{j}\cr
&+g^{3}\bar\psi_{i}\Big[\frac{m^{2}}{2}\frac{\slashed{n}}{n.\partial}(n.G^{a})\frac{1}{n.\partial}
(n.G^{b})\frac{1}{n.\partial}(n.G^{c})\frac{1}{n.\partial}\Big](T^{a}T^{b}T^{c})_{ij}\psi_{j}\cr
&+ig^{4}\bar\psi_{i}\Big[\frac{m^{2}}{2}\frac{\slashed{n}}{n.\partial}(n.G^{a})\frac{1}{n.\partial}
(n.G^{b})\frac{1}{n.\partial}(n.G^{c})\frac{1}{n.\partial}(n.G^{d})\frac{1}{n.\partial}\Big](T^{a}T^{b}T^{c}T^{d})_{ij}\psi_{j}
+\cdots,
\label{eq:2}
\end{align}
where $(a,b,c,d)$ are adjoint indices.
By considering \eqref{eq:2}, we can obtain the relevant Feynman rules, corresponding to the vertices depicted in Fig.~\ref{fig:vertices}, as below
\begin{itemize}
    \item The 3-point function $\langle \bar{\psi} (p) \psi(q) G_\mu^{a} (k)\rangle $
\begin{align}
\Gamma_{\mu}^{a} =-ig\left[\gamma_{\mu}+\frac{m^{2}}{2}\frac{\slashed{n}~n_{\mu}}
{\left(n.p\right)\left(n.q\right)}\right]T^{a}
\label{eq:4-1}
\end{align}
  \item The 4-point function $\langle \bar{\psi} (p) \psi(q) G_\mu^{a} (k_1)G_\nu^{b} (k_2)\rangle $
\begin{equation}
\Gamma_{\mu\nu}^{ab}=-\frac{ig^{2}m^{2}}{2}\frac{\slashed{n}~n_{\mu}n_{\nu}}{(n.p)(n.q)}
\left[\frac{T^{a}T^{b}}{n.(p+k_2)}+\frac{T^{b}T^{a}}{n.(p+k_1)}\right]
\label{eq:4-2}
\end{equation}
  \item The 5-point function $\langle \bar{\psi} (p) \psi(q)G_\mu^{a} (k_1)G_\nu^{b} (k_2)G_\rho^{c} (k_3)\rangle $
\begin{align}
\Gamma_{\mu\nu\rho}^{abc}=-\frac{ig^{3}m^{2}}{2}\frac{\slashed{n}~ n_{\mu}n_{\nu}n_{\rho}}{(n.p)(n.q)}
&\Bigg[\frac{T^{a}T^{b}T^{c}}{n.(p+k_{2}+k_{3})~n.(p+k_{3})}
+\frac{T^{a}T^{c}T^{b}}{n.(p+k_{2}+k_{3})~n.(p+k_{2})}
\nonumber\\
&+
\frac{T^{b}T^{a}T^{c}}{n.(p+k_{1}+k_{3})~n.(p+k_{3})}
+\frac{T^{b}T^{c}T^{a}}{n.(p+k_{1}+k_{3})~n.(p+k_{1})}
\nonumber\\
&+
\frac{T^{c}T^{a}T^{b}}{n.(p+k_{1}+k_{2})~n.(p+k_{2})}
+
\frac{T^{c}T^{b}T^{a}}{n.(p+k_{1}+k_{2})~n.(p+k_{1})}
\Bigg]
\label{eq:4-3}
\end{align}
\item The 6-point function $\langle \bar{\psi} (p) \psi(q)G_\mu^{a} (k_1)G_\nu^{b} (k_2)G_\rho^{c} (k_3)G_{\sigma}^{d}(k_4)\rangle $
\begin{eqnarray}
\Gamma_{\mu \nu \rho \sigma}^{a b c d} &=& -  \frac{i g^4m^2}{2} \frac{\slashed{n}~
n_{\mu} n_{\nu} n_{\rho} n_{\sigma}}{(n.p) (n.q)} \Bigg[\frac{T^a T^b T^c
T^d}{n.(k_2 + k_3 + k_4 + p)~n.(k_3 + k_4 + p)~n.(k_4
+ p)}\nonumber\\
&+& \text{all possible permutations of color indices and momenta}\Bigg]
\end{eqnarray}
\label{eq:4-4}
\end{itemize}
The momenta $p$ and $q$ are related to the fermionic legs, while the momenta $k_i$'s correspond to the gluon legs, which are inward.
Thus, the energy-momentum conservation in each vertex is $p-q+\sum_i k_i=0$.
 It is worth to mention that these Feynman rules are valid in any dimension. However, our analysis will hold in the $(2+1)$-dimensional space-time.

\section{One-loop 2-point function $\langle G_\mu^{a} G_\nu^{b}\rangle$}
\label{sec3}

\begin{figure}[t]
\vspace{-1.2cm}
\includegraphics[height=5.5\baselineskip]{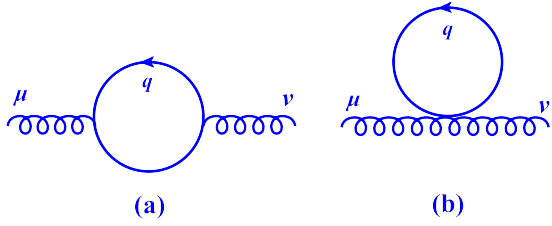}
 \centering\caption{Relevant graphs for the induced $GG$-term.}
\label{fig:oneloop1}
\end{figure}

In order to determine the quadratic part of the gluon's effective action, we shall calculate the two-point function
$\langle G_\mu^{a}G_\nu^{b}\rangle$ at one-loop order, i.e. the term $n=2$ of the series \eqref{eq:1-3}.\footnote{ Here, we remark that the first term of the series \eqref{eq:1-3} corresponds to the gluon's one-point function, which vanishes due to the traceless property of the $SU(3)$ generators.} By considering the Feynman vertices in Fig.~\ref{fig:vertices},
we realize that there are two diagrams contributing to the amplitude of $\langle G_\mu^{a} G_\nu^{b}\rangle$, which are shown in  Fig.~\ref{fig:oneloop1}.
The first diagram (a) arises from the modified quark-gluon vertex in VSR \eqref{eq:4-1}, and the second diagram (b) originates from the new vertex \eqref{eq:4-2}, corresponding to pure VSR effects.

Applying the Feynman rules presented in the previous section, the Feynman expressions of the diagrams (a) and (b) are given by
\begin{align}
\Pi_{\mu\nu}^{ab}\Big|_{\tiny\mbox{\textbf{(a)}}}&=-\int\frac{d^{d}q}{\left(2\pi\right)^{d}}~\textrm{tr}
\Big[\frac{i\left(\slashed{\widetilde{q}}+m_{q}\right)\delta^{i\ell}}{q^{2}-\mu^{2}}
(\Gamma^{a}_{\mu})_{ij}\left(q,p+q\right)\frac{i\left(\slashed{\widetilde{u}} +m_{q}\right)\delta^{jn}}{ u^{2}-\mu^{2}}(\Gamma_{\nu}^{b})_{n\ell}\left(p+q,q\right)\Big],\cr
\Pi_{\mu\nu}^{ab}\Big|_{\tiny\mbox{\textbf{(b)}}}&=-\int\frac{d^{d}q}{\left(2\pi\right)^{d}}
~\textrm{tr}\Big[\frac{i\left(\slashed{\widetilde{q}}+m_{q}\right)\delta^{ij}}{q^{2}-\mu^{2}}
(\Gamma_{\mu\nu}^{ab})_{ij}\left(q,-q,p,-p\right)\Big],
\end{align}
where $u=p+q$. Thus, the full contribution to the two-point function $\langle G_\mu^{a} G_\nu^{b}\rangle$ corresponds to
\begin{equation} \label{eq111}
\Pi_{\mu\nu}^{ab}\Big|_{\tiny\mbox{\textbf{(a+b)}}}=\Pi_{\mu\nu}^{ab}\Big|_{\tiny\mbox{\textbf{(a)}}}
+\Pi_{\mu\nu}^{ab}\Big|_{\tiny\mbox{\textbf{(b)}}}.
\end{equation}
We observe that the computation of the 2-point function \eqref{eq111} is the same as the abelian case. This calculation was already performed in \cite{Bufalo:2020cst}, except for the presence of the SU(3) generators.
Hence, we use the following trace identities in $(2+1)$ dimensions
\begin{align}
\textrm{tr}(\gamma^\mu \gamma^\nu)&=2\eta^{\mu\nu}, \label{tr1}\\
\textrm{tr}(\gamma^\mu \gamma^\nu \gamma^\rho)&=-2i\epsilon^{\mu\nu\rho},  \label{tr2}\\
\textrm{tr}(\gamma^\mu \gamma^\nu \gamma^\rho \gamma^\sigma)&=2(\eta^{\mu\nu}\eta^{\rho\sigma}-\eta^{\mu\rho}\eta^{\nu\sigma}+\eta^{\mu\sigma}\eta^{\nu\rho}),
 \label{tr3}
\end{align}
and the following decomposition formula for the non-local terms,
\begin{equation}
\label{decomposition}
\frac{1}{n.(q+p_i)~n.(q+p_j)}=\frac{1}{n.(p_i-p_j)}\bigg(\frac{1}{n.(q+p_j)}-\frac{1}{n.(q+p_i)}\bigg),
\end{equation}
to keep only one factor $1/(n.q)$ in every VSR term for the momentum integration.
We apply the Feynman parameterization and  the Mandelstam-Leibbrandt prescription \cite{Mandelstam:1982cb,Leibbrandt:1983pj}, as well as the result of \cite{Alfaro:2016pjw} to perform the relevant integrals. Useful momentum integral for this computation can be found in Appendix \ref{apA}.

The Mandelstam-Leibbrandt method \cite{Mandelstam:1982cb,Leibbrandt:1983pj}, used in the momentum integration, introduces a new null vector $\bar{n}$.
However, this new null vector breaks the SIM(1) symmetry.
In order to restore the symmetry in this computation, we follow the proposal presented in \cite{Alfaro:2017umk}.
The main idea is to write $\bar{n}$ as a linear combination of the null vector $n$, present in the VSR theory, and the only one external momentum of the diagram. Thus, requiring some conditions, such as: reality, right scaling $(n,\bar{n})\rightarrow(\lambda n,\lambda^{-1}\bar{n})$ and being dimensionless \cite{Alfaro:2017umk}, we have
\begin{equation}
\bar{n}_\mu=-\frac{p^2}{2(n.p)^2}n_\mu+\frac{p_\mu}{n.p}.
\label{eq:nbar}
\end{equation}
Under these considerations, the one-loop two-point function of the gluon, in the low-energy limit, $p^2\ll m_q^2$,
leads to the induced kinetic term for the Yang-Mills-Chern-Simons action with VSR corrections as
\begin{align}
-i\Pi_{\mu\nu}^{ab}(p)\Big|_{\tiny\mbox{\textbf{(a+b)}}}^{p^{2} \ll m_{q}^{2}}&=\textrm{tr}(T^{a}T^{b})
\bigg\{\frac{g^{2}}{12\pi |m_{q}|}(p_{\mu}p_{\nu}-\eta_{\mu\nu} p^{2})
-\frac{ig^{2}m_q}{4\pi |m_q|}\epsilon_{\mu\nu\rho}p^{\rho}\cr
&+\frac{g^{2}}{16\pi | m_{q}|} \Big(\frac{m^{2}}{m^{2}_{q}}\Big)\Big[\frac{n_{\mu}p_{\nu}+n_{\nu}p_{\mu}}{(n.p)}-\eta_{\mu\nu}
-\frac{n_{\mu}n_{\nu}}{(n.p)^{2}}p^{2}\Big]
p^{2}\cr
&-\frac{ig^{2} m_q}{16\pi |m_q|}  \Big(\frac{m^{2}}{m^{2}_{q}}\Big)\Big[\epsilon_{\mu\nu\rho}\frac{n^{\rho}}{(n.p)}+\epsilon_{\nu\rho\sigma}p^{\rho}
\frac{n_{\mu}n^{\sigma}}{(n.p)^{2}}-\epsilon_{\mu\rho\sigma}p^{\rho}\frac{n_{\nu}n^{\sigma}}{(n.p)^{2}}\Big]
p^{2}\bigg\}.
\label{eq:pi}
\end{align}
The result is the same as the VSR-QED case, except for the overall color factor $\textrm{tr}(T^aT^b)=\frac{1}{2}\delta^{ab}$
 Moreover, \eqref{eq:pi} satisfies the Ward identity as $p^{\mu}\Pi_{\mu\nu}^{ab}(p)=0$ and $p^{\nu}\Pi_{\mu\nu}^{ab}(p)=0$.

As we observe, the first term of \eqref{eq:pi} corresponds to the kinetic term of the Yang-Mills action, whereas the second term corresponds to the kinetic term of the non-abelian Chern-Simons action.
The remaining terms indicate the
(higher-derivative) VSR-nonlocal corrections to the Yang-Mills and Chern-Simons kinetic parts.

Furthermore, the VSR contributions in \eqref{eq:pi} include the ratios of contractions of the null vector with the external momentum as $\frac{n_{\mu}}{(n.p)}$ and $\frac{n_{\mu}n_{\nu}}{(n.p)^{2}}$, which are SIM(1)-invariant. This shows that the SIM(1) invariance is also preserved for the free part of the one-loop gluon's effective action.

\subsection{The tensorial structure of $\Pi_{\mu\nu}^{ab}(p)$}

The tensorial structure for the gluon's 2-point function, found in \eqref{eq:pi}, arises directly from the one-loop analysis.
Hence, it is interesting to obtain this tensorial structure generically, without any explicit computation. In fact, this can be performed only by considering the symmetry content of the amplitude in terms of the Ward identity.
Naturally, in the three-dimensional QCD,  due to the gauge invariance within the VSR setting, we expect to observe such tensorial structure for the gluon's polarization tensor at any order of the loop calculation.

The general form of a rank-2 tensor, built in terms of the following set of three-dimensional vectors and tensors $\left( \eta_{\mu\nu}, p_{\mu}, n_{\mu}, \epsilon_{\mu\nu\rho}\right)$, is given by
\begin{align}
\Pi^{ab}_{\mu \nu}(p)&=\delta^{ab}\Big[\Pi_1 \eta_{\mu\nu}+\Pi_2 p_\mu p_{\nu} +\Pi_3 \frac{n_\mu p_\nu}{(n.p)}
+\Pi_4 \frac{n_\nu p_\mu}{(n.p)}+\Pi_5\frac{n_\mu n_\nu}{(n.p)^2}+\Pi_6 \epsilon_{\mu\nu\rho}p^\rho \nonumber\\
&+\Pi_7 \frac{\epsilon_{\mu\nu\rho} n^\rho}{(n.p)} + \Pi_8\frac{\epsilon_{\mu\rho\sigma}\, p^\rho n^\sigma n_\nu}{(n.p)^2} + \Pi_9 \frac{\epsilon_{\nu\rho\sigma}\, p^\rho n^\sigma n_\mu}{(n.p)^2}\Big],
\label{eq:pi-1}
\end{align}
where the coefficients $\Pi_{i}$'s are not independent quantities, due to the Ward identity.
Now, applying the Ward identity as $p^{\mu} \Pi_{\mu\nu}=0$ and $p^{\nu} \Pi_{\mu\nu}=0$, we get the following constraints
\begin{equation}
\left\{
  \begin{array}{ll}
      \Pi_1 + \Pi_2 p^2+\Pi_3 =0,\\
    \Pi_4 p^2+ \Pi_5 =0,\\
     \Pi_7-\Pi_9 =0,
  \end{array}
\right.
\hspace{2cm}
\left\{
  \begin{array}{ll}
    \Pi_1+ \Pi_2p^2+\Pi_4=0, \\
    \Pi_3p^2+\Pi_5 =0,  \\
   \Pi_7+\Pi_8 =0.
  \end{array}
\right.
\label{eq:cons}
\end{equation}
Here, we remark that since $\Pi^{ab}_{\mu \nu}$ in \eqref{eq:pi-1} is not symmetric under the exchange of $\mu\leftrightarrow\nu$, we have two different constraints sets in \eqref{eq:cons}.
Thus, by solving the relations of \eqref{eq:cons}, the number of independent coefficients in \eqref{eq:pi-1} reduces to four, and hence  the polarization tensor \eqref{eq:pi-1} has the simplified form as below
\begin{align}
\Pi_{\mu\nu}^{ab}(p)&=\delta^{ab}\Big[\Pi_1 \eta_{\mu\nu}+\Pi_2 p_{\mu} p_{\nu} -\frac{(\Pi_1+\Pi_2p^2)}
{(n.p)}\Big(n_{\mu} p_{\nu}+ n_{\nu} p_{\mu}-\frac{ n_{\mu} n_{\nu}}{(n.p)}p^2\Big)\nonumber\\
&+\Pi_6 \epsilon_{\mu\nu\rho}p^\rho +\Pi_7\Big(\epsilon_{\mu\nu\rho}\frac{n^\rho}{(n.p)}
  - \epsilon_{\mu\rho\sigma}\, p^\rho \frac{n_\nu n^\sigma }{(n.p)^2}
 +  \epsilon_{\nu\rho\sigma}\, p^\rho \frac{n_\mu n^\sigma}{(n.p)^2}\Big)\Big].
\label{eq:finalform}
\end{align}
The value of these four coefficients $\left(\Pi_1,\Pi_2,\Pi_6,\Pi_7\right)$ is fully determined by the explicit perturbative computation.
As we see, the tensorial structure of \eqref{eq:pi} is the same as \eqref{eq:finalform}, showing thus the correctness of the tensorial form of the result found in \eqref{eq:pi}.


\section{One-loop 3-point function $\langle G_\mu^{a} G_\nu^{b} G_{\rho}^{c}\rangle$}
\label{sec4}

\begin{figure}[t]
\vspace{-1.2cm}
\includegraphics[height=8\baselineskip]{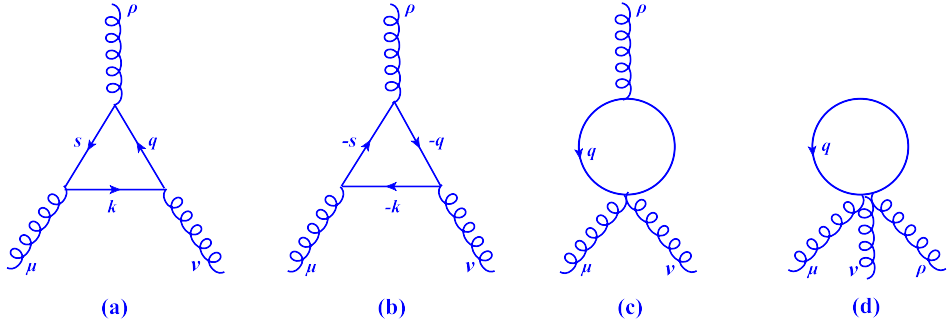}
 \centering\caption{Relevant graphs for the induced $GGG$-term}
\label{fig:threepoint}
\end{figure}

In this section, we present the main purpose of this work which is the computation of the 3-point function $\langle G_\mu^{a} G_\nu^{b} G_{\rho}^{c}\rangle$.
Four graphs contribute to this self-coupling part of the effective action, which are shown in Fig.~\ref{fig:threepoint}.
We observe that the graphs (a) and (b) are the usual diagrams appearing in the standard case (but with VSR modifications), while the graphs (c) and (d) are generated because of the new vertices Eqs.~\eqref{eq:4-2} and \eqref{eq:4-3}, respectively, due to the VSR effects.

As we have mentioned in the introductory section, the non-abelian case presents an important difference with respect to the abelian case, investigated in \cite{Bufalo:2020cst}.
Actually, in \cite{Bufalo:2020cst}, it is proved that the amplitude of the graphs contributing to the photon's 3-point function identically vanishes, and hence, Furry's theorem is satisfied.
In other words, in the usual QED as well as VSR-QED, the vacuum expectation value of any odd number of currents vanishes
 \begin{equation}
\langle \Omega\big|T[j_{\mu_{1}}(x_{1})\cdots j_{\mu_{2n+1}}(x_{2n+1})]\Omega\big|\rangle=0.
\label{eq:furry-abelian}
\end{equation}
This identity can easily be proved by the insertion of the charge conjugation operator $C$ as $C^{\dagger}C=1$, as well as using the fact that $Cj_{\mu}C^{\dagger}=-j_{\mu}$ and the (vacuum state) invariance $C\big|\Omega\rangle=\big|\Omega\rangle$.
On the other hand, in the present case, we can show that the relevant amplitude for the vacuum expectation value of 3 currents  does not vanish, as we would naturally expect in a non-abelian gauge theory.
Actually, the main reason for this non-vanishing amplitude relies on the properties of the symmetric group generators and has nothing to do with VSR effects.
 To clarify this point, we should remind the behaviour of the $SU(3)$ current, $J_{\mu}^{a}=\bar\psi\gamma_{\mu}T^{a}\psi$ under the charge conjugation (C) in a usual $SU(3)$ gauge theory \cite{Peccei:1998jv}, as below
\begin{equation} \label{eq101}
\big(J_{\mu}^{a}\big)^{C}=-\bar\psi(x)\gamma_{\mu}(T^{a})^{t}\psi(x),
\end{equation}
where $(T^{a})^{t}$ indicates the transpose of $T^{a}$. The generators $T^{1},T^{3},T^{4},T^{6}$ and $T^{8}$ are symmetric, whereas $T^{2},T^{5}$ and $T^{7}$ are antisymmetric.
Therefore, we obtain that \eqref{eq101} can be cast as $\big(J_{\mu}^{a}\big)^{C}=-\xi(a)J_{\mu}^{a}(x)$, in which
\begin{equation}
\xi(a)=\left\{
  \begin{array}{ll}
    +1, & a=1, 3, 4, 6, 8; \\
    -1, & a=2, 5, 7.
  \end{array}
\right.
\end{equation}
Hence, in order to have a C-invariant non-abelian gauge theory, it is a sufficient condition that the non-Abelian gauge field transforms as $\big(G_{\mu}^{a}\big)^{C}=-\xi(a)G_{\mu}^{a}(x)$.

Now, if we consider the anti-symmetric generators, i.e. $\xi(a)=\xi(b)=\xi(c)=-1$, we arrive at
\begin{equation}
\langle \Omega\big|T[J_{\mu}^{a}(x)J_{\nu}^{b}(y)J_{\rho}^{c}(z)]\Omega\big|\rangle \neq 0,
\label{eq:furry}
\end{equation}
which is in contrast to the abelian result \eqref{eq:furry-abelian}.
However, in the case of symmetric generators, i.e. $\xi(a)=\xi(b)=\xi(c)=1$, the amplitude \eqref{eq:furry} vanishes.
As a result, Furry's theorem is not fully satisfied in a non-abelian gauge theory.
Similarly, this also happens in a VSR non-abelian gauge theory.
In the following, we shall find the amplitude corresponding to the 3-point function of the gluon perturbatively.
Our result confirms the correctness of the above discussion, in particular Eq.~\eqref{eq:furry}.

The contributing diagrams to this analysis are depicted in Fig.~\ref{fig:threepoint}. The total amplitude of these diagrams will add new self-coupling terms to the kinetic part generated by \eqref{eq:pi}.
In order to have a clearer analysis of these four contributions, we present our computations in separate parts.

\subsection{Contribution of the graphs (a) and (b)}
\label{sec4-1}

First, we will consider the diagrams (a) and (b), which possess the same standard structure. Using the Feynman rules, the amplitude of the diagrams (a) and (b) reads
\begin{align}
\Lambda_{\mu\nu\rho}^{abc}\Big|_{\tiny\mbox{\textbf{(a)}}}&=-g^{3}\int\frac{d^{d}k}{(2\pi)^{d}}
~\textrm{tr}\Big[\frac{(\slashed{\widetilde{k}}+m_{q})}{k^{2}-\mu^{2}}
\Big(\gamma_{\nu}+\frac{m^{2}}{2}\frac{n_{\nu}\slashed{n} }{(n.k)(n.q)}\Big)
\frac{(\slashed{\widetilde{q}}+m_{q})}{q^{2}-\mu^{2}}\cr
&\times
\Big(\gamma_{\rho}+\frac{m^{2}}{2}\frac{n_{\rho}\slashed{n} }{(n.q)(n.s)}\Big)
\frac{(\slashed{\widetilde{s}}+m_{q})}{s^{2}-\mu^{2}}
\Big(\gamma_{\mu}+\frac{m^{2}}{2}\frac{n_{\mu}\slashed{n} }{(n.s)(n.k)}\Big)
\Big]~\textrm{tr}(T^{a}T^{b}T^{c}), \\
\Lambda_{\mu\rho\nu}^{acb}\Big|_{\tiny\mbox{\textbf{(b)}}}&=g^{3}\int\frac{d^{d}k}{(2\pi)^{d}}
~\textrm{tr}\Big[\frac{(\slashed{\widetilde{s}}-m_{q})}{s^{2}-\mu^{2}}
\Big(\gamma_{\rho}+\frac{m^{2}}{2}\frac{n_{\rho}\slashed{n} }{(n.s)(n.q)}\Big)
 \frac{(\slashed{\widetilde{q}}-m_{q})}{q^{2}-\mu^{2}}\cr
&\times
\Big(\gamma_{\nu}+\frac{m^{2}}{2}\frac{n_{\nu}\slashed{n} }{(n.q)(n.k)}\Big) \frac{(\slashed{\widetilde{k}}-m_{q})}{k^{2}-\mu^{2}}
\Big(\gamma_{\mu}+\frac{m^{2}}{2}\frac{n_{\mu}\slashed{n} }{(n.k)(n.s)}\Big)
\Big]~\textrm{tr}(T^{a}T^{c}T^{b}),
\end{align}
where $q=k+p_{2}$ and $s=k-p_{1}$, and the external gluon legs are denoted by incoming momenta $(p_{1\mu}, p_{2\nu}, p_{3\rho})$, satisfying the energy-momentum conservation $p_1+p_2+p_3=0$.
Gathering both contributions, we find that
\begin{align}
\label{diagabsum}
\Lambda_{\mu\nu\rho}^{abc}\Big|_{\tiny\mbox{\textbf{(a+b)}}}=-g^{3}\int\frac{d^{d}k}{(2\pi)^{d}}
\frac{N_{\mu\nu\rho}^{\alpha\beta\lambda}}{(k^{2}-\mu^{2})(q^{2}-\mu^{2})(s^{2}-\mu^{2})}
\Big[\textrm{tr}(T^{a}T^{b}T^{c}){\pazocal{A}}_{\alpha\beta\lambda}+\textrm{tr}(T^{a}T^{c}T^{b}){\pazocal{B}}_{\beta\alpha\lambda}\Big],
\end{align}
where
\begin{align}
N_{\mu\nu\rho}^{\alpha\beta\lambda}&=\bigg[\eta_{\nu}^{\alpha}+\frac{m^{2}}{2}\frac{n^{\alpha} n_{\nu}}{(n.k)(n.q)}\bigg]\bigg[\eta_{\rho}^{\beta}+\frac{m^{2}}{2}\frac{n^{\beta} n_{\rho}}{(n.q)(n.s)}\bigg]
\bigg[\eta_{\mu}^{\lambda}+\frac{m^{2}}{2}\frac{n^{\lambda} n_{\mu}}{(n.s)(n.k)}\bigg],\\
{\pazocal{A}}_{\alpha\beta\lambda}&=\textrm{tr}\Big[(\slashed{\widetilde{k}}+m_{q})
\gamma_{\alpha}
(\slashed{\widetilde{q}}+m_{q})
\gamma_{\beta}
(\slashed{\widetilde{s}}+m_{q})
\gamma_{\lambda}
\Big],\\
{\pazocal{B}}_{\beta\alpha\lambda}&=-\textrm{tr}\Big[(\slashed{\widetilde{s}}-m_{q})
\gamma_{\beta}
(\slashed{\widetilde{q}}-m_{q})
\gamma_{\alpha}
(\slashed{\widetilde{k}}-m_{q})
\gamma_{\lambda}
\Big].
\end{align}
The first point about \eqref{diagabsum} is that the presence of the trace of the SU(3) generators makes a difference with respect to the abelian case, where the sum of both diagrams vanishes. In this case, this contribution is not zero since $T^bT^c\neq T^cT^b$, giving us the induced self-coupling terms.

From the identity $C^{-1}\gamma^{\mu}C=-(\gamma^{\mu})^{T}$, we can show that for any number of gamma matrices, the following relation
\begin{equation}
\textrm{tr}\Big(\gamma^{\mu_{1}}\gamma^{\mu_{2}}\ldots\gamma^{\mu_{n-1}}\gamma^{\mu_{n}}\Big)=
(-1)^{n}\textrm{tr}\Big(\gamma^{\mu_{n}}\gamma^{\mu_{n-1}}\ldots\gamma^{\mu_{2}}\gamma^{\mu_{1}}\Big),
\end{equation}
is satisfied. Applying this identity, we obtain that
\begin{equation} \label{idab}
{\pazocal{B}}_{\beta\alpha\lambda}=-{\pazocal{A}}_{\alpha\beta\lambda}.
\end{equation}
Moreover, the trace of three matrices $T^a$ can be expressed as
\begin{equation}
\label{tract}
\textrm{tr}[T^{a}T^{b}T^{c}]=\frac{1}{4}(d^{abc}+i f^{abc}),
\end{equation}
where $d^{abc}$ is a totally symmetric tensor, defined as $d^{abc}=2\textrm{tr}[\{T^a,T^b\}T^c]$, and $f^{abc}$ are the structure constants of the group.
Hence, by considering Eqs.~\eqref{idab} and \eqref{tract}, we find that the coefficient of the symmetric part, $d^{abc}$, in \eqref{diagabsum} vanishes and we are left only with the anti-symmetric part, $f^{abc}$, as below
\begin{equation}
\label{abdiag}
\Lambda_{\mu\nu\rho}^{abc}\Big|_{\tiny\mbox{\textbf{(a+b)}}}=-\frac{ig^{3}}{2}f^{abc}\int\frac{d^{d}k}{(2\pi)^{d}}
\frac{N_{\mu\nu\rho}^{\alpha\beta\lambda}}{(k^{2}-\mu^{2})(q^{2}-\mu^{2})(s^{2}-\mu^{2})} ~{\pazocal{A}}_{\alpha\beta\lambda}.
\end{equation}

Now, we can evaluate the numerator of \eqref{abdiag} and make use of the trace identities, \eqref{tr1}-\eqref{tr3}, to simplify the expression $N_{\mu\nu\rho}^{\alpha\beta\lambda}{\pazocal{A}}_{\alpha\beta\lambda}$.
Next, we make use the decomposition formula \eqref{decomposition} to have only one nonlocal term $1/(n.k_i)$ in the VSR parts.
 After applying the Feynman parametrization  and change of variables, we are faced with a huge number of integrals with the nonlocal factor $1/(n.k_i)$ appearing in the computation.
In order to solve them, we use the result listed in Appendix \ref{apA}.

In the next step, since we are interested in finding the low-energy effective action for the gluon, we take the limit $p_i^2\ll m_{q}^2$. From these considerations, we finally obtain the total amplitude of these two diagrams as below
\begin{equation}
\Lambda_{\mu\nu\rho}^{abc}\Big|_{\tiny\mbox{\textbf{(a+b)}}}^{p_{i}^{2}\ll m_{q}^{2}}=\Lambda_{\mu\nu\rho}^{abc}\Big|_{\small\mbox{(st)}}+\Lambda_{\mu\nu\rho}^{abc}\Big|_{\small\mbox{(vsr)}},
\label{two-parts}
\end{equation}
where we have separated the result into two parts: The first one includes the standard terms, whereas the second one contains the terms proportional to the powers of $m$, which are purely VSR effects.

The explicit form of $\Lambda^{\mu\nu\rho}_{abc}\Big|_{\small\mbox{(st)}}$ is given by
\begin{align}
\label{lst}
\Lambda_{\mu\nu\rho}^{abc}\Big|_{\small\mbox{(st)}}=
\frac{ig^3 m_q f^{a b c} }{8 \pi |m_q|} \varepsilon_{\mu \nu \rho}
+ \frac{g^3f^{a b c} }{24 \pi|m_q|} \Big[(p_{1} - p_{2})_{\rho} \eta_{\mu \nu} -
(p_{2} + 2 p_{1})_{\nu} \eta_{\mu \rho} + (p_1 + 2 p_2)_{\mu} \eta_{\nu
\rho}\Big].
\end{align}
The first term of \eqref{lst} corresponds to the interaction term of the non-abelian Chern-Simons action, with odd parity. The second term is identified as the Yang-Mills cubic coupling, with even parity.
Gathering this parity-odd self-coupling in \eqref{lst} and also the kinetic term of the ordinary Chern-Simons in \eqref{eq:pi}, we find the corresponding induced effective action for the gluons with odd parity, named as the ordinary (non-VSR) non-abelian Chern-Simons action ($m\rightarrow 0$)
\begin{equation}
\label{CS-action}
\Gamma^{\tiny\mbox{CS}}_{\tiny\mbox{eff}}[G]=\frac{g^2}{8\pi}\frac{m_q}{|m_q|}\int d^{3}x~\varepsilon^{\mu\nu\rho}\Big[G_{\mu}^{a}(x)\partial_{\nu}G_{\rho}^{a}(x)+\frac{ig}{3}
f^{abc}G_{\mu}^{a}(x)G_{\nu}^{b}(x)G_{\rho}^{c}(x)\Big],
\end{equation}
or equivalently
\begin{equation}
\Gamma^{\tiny\mbox{CS}}_{\tiny\mbox{eff}}[G]=\frac{g^2}{4\pi}\frac{m_q}{|m_q|}\int d^{3}x~\varepsilon^{\mu\nu\rho}\textrm{tr}\Big[G_{\mu}(x)\partial_{\nu}G_{\rho}(x)+\frac{2g}{3}
G_{\mu}(x)G_{\nu}(x)G_{\rho}(x)\Big],
\end{equation}
which is invariant under the infinitesimal non-abelian gauge transformations \eqref{eq105}, with $\kappa\rightarrow 0$.

The action $\Gamma^{\tiny\mbox{CS}}_{\tiny\mbox{eff}}$ \eqref{CS-action} is known to be a topological theory.
A remarkable property of this kind of models is that the corresponding energy-momentum tensor
\begin{equation}
T_{\mu\nu}^{\tiny\mbox{CS}}=\frac{2}{\sqrt{-\det\eta}}\frac{\delta\Gamma^{\tiny\mbox{CS}}_{\tiny\mbox{eff}}}{\delta\eta^{\mu\nu}},
\end{equation}
is zero due to the metric independence of $\Gamma^{\tiny\mbox{CS}}_{\tiny\mbox{eff}}$.
Therefore, the relevant Hamiltonian vanishes, $H^{\tiny\mbox{CS}}=0$.

Moreover, the second term in the first line of \eqref{lst} can be rewritten by applying the energy-momentum conservation $p_1+p_2+p_3=0$ as below
\begin{equation}
\frac{g^3f^{a b c}}{24 \pi|m_q|} \Big[(p_1 - p_2)_{\rho} \eta_{\mu \nu} +(p_2 - p_3)_{\mu} \eta_{\nu \rho}
+(p_3 - p_1)_{\nu} \eta_{\mu \rho}\Big],
\label{eq:3-gluon-vertex}
\end{equation}
which corresponds to the usual 3-gluon interaction term with even parity in the Yang-Mills action.

Now, about the VSR part of \eqref{two-parts}, we see that it contains a large number of terms that can be decomposed depending on the powers of VSR parameter $m$. Thus, we have
\begin{equation}
\Lambda_{\mu\nu\rho}^{abc}\Big|_{\small\mbox{(vsr)}}=\Lambda_{\mu\nu\rho}^{abc}\Big|_{\small\mbox{(vsr)}}^{(2)}
+\Lambda_{\mu\nu\rho}^{abc}\Big|_{\small\mbox{(vsr)}}^{(4)}+\Lambda_{\mu\nu\rho}^{abc}\Big|_{\small\mbox{(vsr)}}^{(6)},
\label{eq:vsr-cont}
\end{equation}
where the number in the index $j$ in the function $\Lambda_{\mu\nu\rho}^{abc}\Big|_{\small\mbox{(vsr)}}^{(j)}$ indicates the power of the VSR parameter $m$.
Since, the parameter $m$ is expected to be very small, we shall consider the leading contribution in powers of $m^2/m_q^2$, thus neglecting the terms proportional to $m^4$ and $m^6$ in \eqref{eq:vsr-cont}. The explicit analysis of the $m^2$ part yields the following expression
\begin{align}
\Lambda_{\mu\nu\rho}^{abc}\Big|_{\small\mbox{(vsr)}}^{(2)}&=\frac{i g^3f^{abc}}{64\pi m_q^2}\left(\frac{m^2}{m_{q}^2}\right)\int_0^1 dx \int_0^{1-x} dy~\pazocal{M}_{\mu\nu\rho}(x,y,p_1,p_2,n),
\label{eq:lambdavsr}
\end{align}
where the tensor $\pazocal{M}_{\mu\nu\rho}$ contains a huge number of terms, depending on the external momenta $(p_1,p_2)$, Feynman parameters $(x,y)$ and the VSR null vector $n_\mu$.

\subsection{Contribution of the graph (c)}
\label{sec4-2}

 Let us now analyze the diagram (c) of Fig.~\ref{fig:threepoint}. Hence, the corresponding Feynman expression is given by
\begin{align}
\Lambda_{\mu\nu\rho}^{abc}\Big|_{\tiny\mbox{\textbf{(c)}}}&= -\frac{g^{3}m^{2}}{2}n_{\mu}n_{\nu}\int\frac{d^{d}q}
{(2\pi)^{d}}\frac{1}{(n.q)\,
(n.u)}~\textrm{tr}\bigg[\Big(\gamma_{\rho}+\frac{m^{2}}{2}\frac{n_{\rho}\slashed{n}}{(n.q )\,
(n.u)}\Big)T_{ij}^{a}\Big(\frac{\slashed{\widetilde{u}}+m_{q}}{u^{2}-\mu^{2}}\Big)\delta^{il}\slashed{n}\cr
&  \times\Big(\frac{1}{n.(p_{2}+u)}(T^{c}T^{b})_{kl}+\frac{1}{n.(p_{3}+u)}(T^{b}T^{c})_{kl}\Big)
\Big(\frac{\slashed{\widetilde{q}}+m_{q}}{q^{2}-\mu^{2}}\Big)\delta^{kj}\bigg].
\end{align}
Here, we recall that the momenta $p_i$'s are incoming and $u=p_{1}+q$.
Using the decomposition \eqref{tract}, we realize that the above expression can be divided into two parts: symmetric and anti-symmetric with respect to the adjoint indices, that is
\begin{equation}
\Lambda_{\mu\nu\rho}^{abc}\Big|_{\tiny\mbox{\textbf{(c)}}}=f^{abc}{\pazocal{J}}_{\mu\nu\rho}^{\tiny\mbox{\textbf{(c)}}}+ d^{abc}{\pazocal{K}}_{\mu\nu\rho}^{\tiny\mbox{\textbf{(c)}}}.
\label{eq:general-structure}
\end{equation}
The coefficient of the symmetric part, ${\pazocal{K}}_{\mu\nu\rho}^{\tiny\mbox{\textbf{(c)}}}$, has exactly the same structure of the abelian case and it vanishes \cite{Bufalo:2020cst}.
Therefore, we focus on the computation of the antisymmetric part, ${\pazocal{J}}_{\mu\nu\rho}^{\tiny\mbox{\textbf{(c)}}}$, which does not have a counter-part in the abelian case \cite{Bufalo:2020cst}.
By making use of the identity \eqref{decomposition} and the integral of Appendix \ref{apA}, we arrive at the following expression at the low-energy limit $p_i^2\ll m_{q}^2$
\begin{align}
\Lambda_{\mu\nu\rho}^{abc}\Big|_{\tiny\mbox{\textbf{(c)}}}^{p_{i}^{2}\ll m_{q}^{2}}&= -\frac{ m^2 g^3f^{a b c}
n_{\mu}n_{\nu} }{32\pi |m_q|(n.p_{1})}\Bigg \lbrace \frac{p_{1}^{2}p_{1\rho}}{3 m_{q}^{2}~n.(p_{1}+p_{2})}
+ \frac{1}{2}\Big\{ \frac{5 n_{\rho}(\bar{n}.p_{1})n.(p_1+2p_2)}{(n.p_2)n.(p_1 + p_2)}-\frac{n.p_1}{(n.p_2)n.(p_1 + p_2)}\cr
&  \times  \Big[5 n_{\rho}\bar{n}.(p_{1}+2p_{2})
+\frac{2}{m_{q}^{2}}\bar{n}.(p_{1}+2p_{2})\Big((p_{1}.p_{2}+ p_{2}^{2})n_{\rho}-(n.p_{2}) \left(p_{1}+2p_{2}\right)_{\rho}-(n.p_{1}) p_{2\rho}\Big)\Big] \Big\}  \Bigg\rbrace\cr
& + \frac{im^2 g^3 f^{abc}}{32 \pi m_{q}^{2}}~\varepsilon_{\rho \alpha \beta} p_{1}^{\alpha} n^{\beta}
n_{\mu}n_{\nu} \Big[\frac{1}{(n.p_{1}) n.(p_{1}+p_{2})}\left(\frac{p_{1}^{2}}{n.p_{1}} + \frac{p_{2}^{2}}{n. p_{2}}\right)-\frac{p_{2}^{2}}{(n.p_{1}) (n.p_{2})^{2}}\Big].
\end{align}
We now replace $\bar{n}$ in terms of the expression \eqref{eq:nbar}. Thus, we get
\begin{align}
\Lambda_{\mu\nu\rho}^{abc}\Big|_{\tiny\mbox{\textbf{(c)}}}^{p_{i}^{2}\ll m_{q}^{2}}&=-
  \frac{ m^2 g^3  f^{a b c}n_{\mu}n_{\nu}  }{32\pi |m_q|(n.p_{1})}\Bigg \lbrace
\frac{p_{1}^{2}p_{1\rho}}{3 m_{q}^{2}~n.(p_{1}+p_{2})} + \frac{1}{2} \Big\lbrace \frac{5 n_{\rho} p_{1}^{2}}{(n.p_{1})
n.(p_1 + p_2)}
 -\frac{n.p_1}{(n.p_2)n.(p_1 + p_2)} \cr
 &\times
\Big[\frac{5 n_{\rho}p_{2}^{2}}{n.p_{2}}
 +\frac{2}{m_{q}^{2}}\Big(\frac{p_{1}^{2}}{2(n.p_{1})}+\frac{p_{2}^{2}}{n.p_{2}}\Big)\Big((p_{1}.p_{2}+p_{2}^{2})n_{\rho}
 -(n.p_{2}) (p_{1}+2p_{2})_{\rho}-(n.p_{1}) p_{2\rho}\Big)\Big]\Big\rbrace \Bigg\rbrace
\cr
&  + \frac{i m^2  g^3 f^{abc}}{32 \pi m_{q}^{2}}~\varepsilon_{\rho \alpha \beta} p^{\alpha}_{1} n^{\beta}
n_{\mu}n_{\nu} \Big[\frac{1}{(n.p_{1}) n.(p_{1}+p_{2})}\left(\frac{p_{1}^{2}}{n.p_{1}} + \frac{p_{2}^{2}}{n. p_{2}}\right)-\frac{p_{2}^{2}}{(n.p_{1}) (n.p_{2})^{2}}\Big].
\label{eq:final-c}
\end{align}
We observe that the above result indicates the presence of VSR corrections to the 3-gluon vertex. The terms in the first two lines of \eqref{eq:final-c} describe the VSR modifications to the 3-gluon vertex in Yang-Mills theory,
while the third line of \eqref{eq:final-c} is the VSR modifications to the 3-gluon vertex in non-abelian Chern-Simons theory.
Another important aspect of this correction is that, as it happened in the 2-point function \eqref{eq:pi}, the VSR contributions are all higher-derivative terms.
However, in the non-VSR limit ($m\to 0$), these contributions vanish, as we expected.

\subsection{Contribution of the graph (d)}
\label{sec4-3}

On the last contribution, coming from graph (d), by using \eqref{eq:4-3}, the relevant amplitude is written as
\begin{align} \label{eq222}
\Lambda_{\mu\nu\rho}^{abc}\Big|_{\tiny\mbox{\textbf{(d)}}}&=\frac{im^{2}g^{3}}{2}n_{\mu}n_{\nu}n_{\rho}\int\frac{d^{d}q}{(2\pi)^{d}}
~\textrm{tr}\Big[\frac{i(\slashed{\widetilde{q}}+m_{q})}{q^{2}-\mu^{2}}
\frac{\slashed{n}}{(n.q)(n.q)}
\cr
&\times \Big(\frac{1}{n.(q+p_{2}+p_{3})~n.(q+p_{3})}T^{a}T^{b}T^{c}
+\frac{1}{n.(q+p_{2}+p_{3})~n.(q+p_{2})}T^{a}T^{c}T^{b}
\cr
&+ \frac{1}{n.(q+p_{1}+p_{3})~n.(q+p_{3})}T^{b}T^{a}T^{c}
+\frac{1}{n.(q+p_{1}+p_{3})~n.(q+p_{1})}T^{b}T^{c}T^{a}
\cr
&+ \frac{1}{n.(q+p_{1}+p_{2})~n.(q+p_{2})}T^{c}T^{a}T^{b}
+ \frac{1}{n.(q+p_{1}+p_{2})~n.(q+p_{1})}T^{c}T^{b}T^{a}
\Big)\Big].
\end{align}
Moreover, by applying the energy-momentum conservation relation $p_1+p_2+p_3=0$ and the trace identity of the $\gamma$ matrices in \eqref{eq222}, we find
\begin{align}
\Lambda_{\mu\nu\rho}^{abc}\Big|_{\tiny\mbox{\textbf{(d)}}}&=-m^{2}g^{3}n_{\mu}n_{\nu}n_{\rho}\int\frac{d^{d}q}{(2\pi)^{d}}
\frac{1}{(q^{2}-\mu^{2})}
\frac{1}{(n.q)} \cr
&\times
\Big[\frac{1}{n.(q-p_{1})~n.(q+p_{3})}~\textrm{tr}(T^{a}T^{b}T^{c})
+ \frac{1}{n.(q-p_{1})~n.(q+p_{2})}~\textrm{tr}(T^{a}T^{c}T^{b})
\cr
&+ \frac{1}{n.(q-p_{2})~n.(q+p_{3})}~\textrm{tr}(T^{b}T^{a}T^{c})
+\frac{1}{n.(q-p_{2})~n.(q+p_{1})}~\textrm{tr}(T^{b}T^{c}T^{a})
\cr
&+ \frac{1}{n.(q-p_{3})~n.(q+p_{2})}~\textrm{tr}(T^{c}T^{a}T^{b})
+ \frac{1}{n.(q-p_{3})~n.(q+p_{1})}~\textrm{tr}(T^{c}T^{b}T^{a})
\Big].
\end{align}
 By means of the decomposition formula \eqref{decomposition} and some suitable change of variables, after a long algebraic manipulation, we find the following structure for this amplitude
\begin{equation}
\Lambda_{\mu\nu\rho}^{abc}\Big|_{\tiny\mbox{\textbf{(d)}}}=f^{abc}{\pazocal{J}}_{\mu\nu\rho}^{\tiny\mbox{\textbf{(d)}}}+ d^{abc}{\pazocal{K}}_{\mu\nu\rho}^{\tiny\mbox{\textbf{(d)}}},
\end{equation}
which is similar to the result of graph (c) in \eqref{eq:general-structure}.
It can be easily shown that the expression of ${\pazocal{K}}_{\mu\nu\rho}^{\tiny\mbox{\textbf{(d)}}}$ is the same as the abelian model in VSR, and thus, it identically vanishes \cite{Bufalo:2020cst}.
Hence, we are left with the antisymmetric part ${\pazocal{J}}_{\mu\nu\rho}^{\tiny\mbox{\textbf{(d)}}}$ that leads to the following result
\begin{equation}
\Lambda_{\mu\nu\rho}^{abc}\bigg|_{\tiny\mbox{\textbf{(d)}}}=-\frac{i}{4}m^{2}g^{3}f^{abc}n_{\mu}n_{\nu}n_{\rho}\pazocal{H},
\end{equation}
where
\begin{align}
{\pazocal{H}}=- \frac{2}{(n.p_{1})(n.p_{2})(n.p_{3})}\Big[&
n.(p_{2}-p_{1})\int\frac{d^{d}q}{(2\pi)^{d}}\frac{1}{((q+p_{3})^{2}-\mu^{2})(n.q)}\nonumber\\
+&n.(p_{1}-p_{3})\int\frac{d^{d}q}{(2\pi)^{d}}\frac{1}{((q+p_{2})^{2}-\mu^{2})(n.q)}\nonumber\\
+&n.(p_{3}-p_{2})\int\frac{d^{d}q}{(2\pi)^{d}}\frac{1}{((q+p_{1})^{2}-\mu^{2})(n.q)}
\Big].
\end{align}
Considering the integral of the Appendix \ref{apA} and taking the low-energy limit $p_i^2\ll m_{q}^2$, we have
\begin{equation}
\Lambda_{\mu\nu\rho}^{abc}\bigg|_{\tiny\mbox{\textbf{(d)}}}^{p_{i}^{2}\ll m_{q}^{2}}=\frac{m^{2}g^{3}}{8\pi |m_q|}
\frac{f^{abc}n_{\mu}n_{\nu}n_{\rho}}{(n.p_{1})(n.p_{2})(n.p_{3})}\Big[
n.(p_{2}-p_{1})(\bar n.p_{3})
+n.(p_{1}-p_{3})(\bar n.p_{2})
+n.(p_{3}-p_{2})(\bar n.p_{1})
\Big].
\end{equation}
Furthermore, replacing the expression of $\bar{n}$, we finally obtain
\begin{equation}
\Lambda_{\mu\nu\rho}^{abc}\bigg|_{\tiny\mbox{\textbf{(d)}}}^{p_{i}^{2}\ll m_{q}^{2}}=\frac{m^{2}g^{3}}{16\pi |m_q|}
\frac{f^{abc}n_{\mu}n_{\nu}n_{\rho}}{(n.p_{1})(n.p_{2})(n.p_{3})}\Bigg[
\frac{n.(p_{2}-p_{1})p^{2}_{3}}{(n.p_{3})}
+\frac{n.(p_{1}-p_{3})p^{2}_{2}}{(n.p_{2})}
+\frac{n.(p_{3}-p_{2})p^{2}_{1}}{(n.p_{1})}
\Bigg].
\label{eq:final-d}
\end{equation}
This expression, similarly to graph (c), demonstrates the pure VSR modifications to the 3-gluon vertex.
However, this term includes only VSR corrections to the 3-gluon vertex in Yang-Mills theory, without any VSR correction to the non-abelian Chern-Simons action.

 As a result of the analysis performed in this section, we realize that the total amplitude of the 3-point function $\langle G_\mu^{a} G_\nu^{b} G_{\rho}^{c}\rangle$, according to the outcomes in Eqs.~\eqref{two-parts},~\eqref{eq:final-c} and \eqref{eq:final-d}, does not vanish
\begin{equation}
\Lambda_{\mu\nu\rho}^{abc}\bigg|_{\tiny\mbox{\textbf{Total}}}=
\Lambda_{\mu\nu\rho}^{abc}\bigg|_{\tiny\mbox{\textbf{(a+b)}}}
+\Lambda_{\mu\nu\rho}^{abc}\bigg|_{\tiny\mbox{\textbf{(c)}}}
+
\Lambda_{\mu\nu\rho}^{abc}\bigg|_{\tiny\mbox{\textbf{(d)}}}\neq 0.
\end{equation}
Nonetheless, this non-zero result was indeed expected based on the aforementioned discussion in the first part of Sec.~\ref{sec4}.
 Hence, Furry's theorem is not satisfied, unlike the abelian case studied in \cite{Bufalo:2020cst}.
Actually, the total amplitude up to the order $m^2$ can be written as
\begin{equation}
\Lambda_{\mu\nu\rho}^{abc}\bigg|_{\tiny\mbox{\textbf{Total}}}=g^{3}f^{abc}
\Big(\pazocal{Q}_{\mu\nu\rho}+m^{2}\pazocal{R}_{\mu\nu\rho}\Big).
\label{eq:totalamp}
\end{equation}
Here, the tensors $\pazocal{Q}_{\mu\nu\rho}$ and $\pazocal{R}_{\mu\nu\rho}$ represent the ordinary and VSR parts, respectively, and their explicit expressions are determined using the relations \eqref{lst}, \eqref{eq:lambdavsr}, \eqref{eq:final-c} and \eqref{eq:final-d}.
Moreover, we observe that both parts in the amplitude \eqref{eq:totalamp} are totally antisymmetric with respect to the gauge group indices.
This property is rather expected since \eqref{eq:totalamp}, in fact,  generates the induced three gluon vertex \eqref{lst} as well as its VSR modifications, Eqs.~\eqref{eq:lambdavsr},\eqref{eq:final-c} and  \eqref{eq:final-d}, in the three-dimensional spacetime.

Furthermore, similar to the previous discussion after \eqref{eq:pi}, we see that the VSR corrections in Eqs.~\eqref{eq:lambdavsr}, \eqref{eq:final-c} and \eqref{eq:final-d} have an explicit SIM(1)-invariant form.
It demonstrates the preservation of the SIM(1) invariance at one-loop level for the interacting part of the gluon's effective action.
In the case of the non-VSR limit ($m\rightarrow 0$), we obtain the Lorentz-invariant terms in the free and interacting parts of the gluon's effective action.
 In both cases, the tree-level (classical-level) spacetime symmetry also exists at loop-level (quantum-level) and is not anomalous.

\section{One-loop 4-point function $\langle G_\mu^{a} G_\nu^{b} G_{\rho}^{c}G_{\sigma}^{d}\rangle$}
\label{sec5}

\begin{figure}[t]
\vspace{-1.2cm}
\includegraphics[height=8.2\baselineskip]{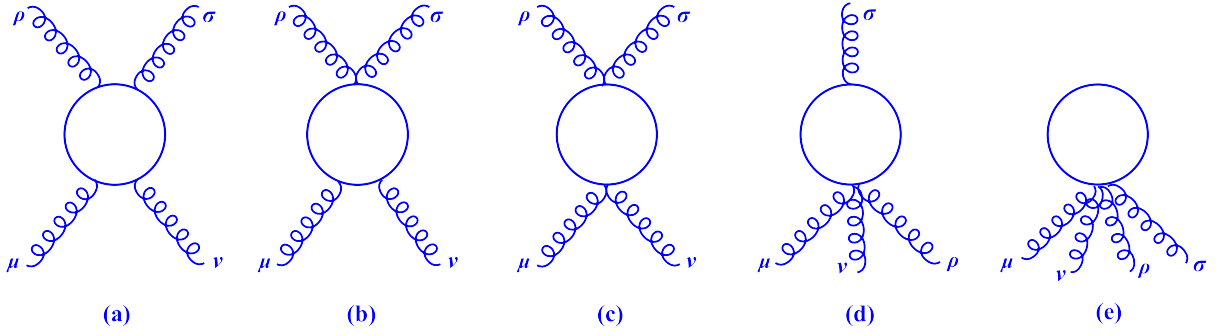}
 \centering\caption{Relevant graphs for the induced $GGGG$-term}
\label{fig:fourpoint}
\end{figure}

Although the main analysis of this work was presented in the previous sections, here we wish to point out to
the graphs that describe the gluon's 4-point function at one-loop level.
There are five diagrams contributing to $\langle G_\mu^{a} G_\nu^{b} G_{\rho}^{c}G_{\sigma}^{d}\rangle$ at order $g^{4}$, depicted in Fig.~\ref{fig:fourpoint}.
Graph (a) includes both VSR and non-VSR contributions, while the other graphs (b), (c), (d) and (e) contain pure VSR effects.
Moreover, we notice that the relevant loop integrals of these five diagrams in the VSR setup generate a huge number of terms, which are more complicated than the analysis of the one-loop 3-point function presented in Sec.~\ref{sec4}.

Since we are interested in obtaining the 4-gluon vertex of the ordinary Yang-Mills action, we focus only on the first graph, which is non-zero in the case of $m\rightarrow 0$. The Feynman amplitude of this diagram is given by
 \begin{align}
\Omega_{\mu\nu\rho\sigma}^{abcd}\Big|_{\tiny\mbox{\textbf{(a)}}}^{(1)}&=-g^{4}\int\frac{d^{d}k}{(2\pi)^{d}}
~\textrm{tr}\Big[\frac{(\slashed{\widetilde{k}}+m_{q})}{k^{2}-\mu^{2}}
\Big(\gamma_{\nu}+\frac{m^{2}}{2}\frac{n_{\nu}\slashed{n} }{(n.k)(n.q)}\Big)
\frac{(\slashed{\widetilde{q}}+m_{q})}{q^{2}-\mu^{2}}
 \Big(\gamma_{\sigma}+\frac{m^{2}}{2}\frac{n_{\sigma}\slashed{n} }{(n.q)(n.u)}\Big)\cr
 &\times\frac{(\slashed{\widetilde{u}}+m_{q})}{u^{2}-\mu^{2}}
\Big(\gamma_{\rho}+\frac{m^{2}}{2}\frac{n_{\rho}\slashed{n} }{(n.u)(n.s)}\Big)
\frac{(\slashed{\widetilde{s}}+m_{q})}{s^{2}-\mu^{2}}
\Big(\gamma_{\mu}+\frac{m^{2}}{2}\frac{n_{\mu}\slashed{n} }{(n.s)(n.k)}\Big)
\Big]~\textrm{tr}(T^{a}T^{b}T^{c}T^{d}),
\label{eq:4-point-first}
\end{align}
where $q=k+p_{2}$, $u=k-p_{1}-p_{3}$ and $s=k-p_{1}$, and the external gluon legs are denoted by incoming momenta $(p_{1\mu}, p_{2\nu}, p_{3\rho},p_{4\sigma})$, satisfying the energy-momentum conservation $p_1+p_2+p_3+p_4=0$.
Analogously to the analysis of the section \ref{sec4-1}, we first simplify the relation \eqref{eq:4-point-first} as
\begin{align}
\Omega_{\mu\nu\rho\sigma}^{abcd}\Big|_{\tiny\mbox{\textbf{(a)}}}^{(1)}=-g^{4}\int\frac{d^{d}k}{(2\pi)^{d}}
\frac{N_{\mu\nu\rho\sigma}^{\alpha\beta\lambda\xi}~{\pazocal{C}}_{\alpha\beta\lambda\xi}}{(k^{2}-\mu^{2})(q^{2}-\mu^{2})(s^{2}-\mu^{2})}
~\textrm{tr}(T^{a}T^{b}T^{c}T^{d})
,
\end{align}
where
\begin{align}
N_{\mu\nu\rho\sigma}^{\alpha\beta\lambda\xi}&=\bigg[\eta_{\nu}^{\alpha}+\frac{m^{2}}{2}\frac{n^{\alpha} n_{\nu}}{(n.k)(n.q)}\bigg]\bigg[\eta_{\sigma}^{\beta}+\frac{m^{2}}{2}\frac{n^{\beta} n_{\sigma}}{(n.q)(n.u)}\bigg]
\bigg[\eta_{\rho}^{\lambda}+\frac{m^{2}}{2}\frac{n^{\lambda} n_{\rho}}{(n.u)(n.s)}\bigg]
\bigg[\eta_{\mu}^{\xi}+\frac{m^{2}}{2}\frac{n^{\xi} n_{\mu}}{(n.s)(n.k)}\bigg]
,\nonumber\\
{\pazocal{C}}_{\alpha\beta\lambda\xi}&=\textrm{tr}\Big[(\slashed{\widetilde{k}}+m_{q})
\gamma_{\alpha}
(\slashed{\widetilde{q}}+m_{q})
\gamma_{\beta}
(\slashed{\widetilde{u}}+m_{q})
\gamma_{\lambda}
(\slashed{\widetilde{s}}+m_{q})
\gamma_{\xi}
\Big].
\end{align}
Then, after taking the loop integral and using the following identity for SU(3) generators
\begin{eqnarray}
 \textrm{tr}(T^{a}T^{b}T^{c}T^{d})=\frac{1}{12}\delta^{ab}\delta^{cd}+\frac{1}{8}(d^{abe}+if^{abe})(d^{cde}+if^{cde}),
\end{eqnarray}
 as well as performing all of 24 permutations of the external legs, we arrive at
\begin{eqnarray}
\sum\limits_{i=1}^{24}\Omega^{abcd}_{\mu\nu\rho\sigma}\Big|_{\tiny\mbox{\textbf{(a)}}}^{(i)}
=\Theta^{abcd}_{\mu\nu\rho\sigma}(p_1,p_2,p_3)\Big|_{\tiny\mbox{\textbf{(a)}}}
+m^{2}\Upsilon^{abcd}_{\mu\nu\rho\sigma}(p_1,p_2,p_3,n)\Big|_{\tiny\mbox{\textbf{(a)}}}.
\label{eq:4-point}
\end{eqnarray}
The first term of \eqref{eq:4-point} indicates the ordinary contribution to the 4-gluon interaction, that in the low-energy limit $p^{2}_i\ll m^{2}_{q}$ is given by
\begin{align}
\Theta^{abcd}_{\mu\nu\rho\sigma}\Big|_{\tiny\mbox{\textbf{(a)}}}=- \frac{ig^{4}}{8\pi |m_q|}\Big[f^{abe}f^{cde}(\eta_{\mu\rho}\eta_{\nu\sigma}-\eta_{\mu\sigma}\eta_{\nu\rho})
+f^{ace}f^{bde}(\eta_{\mu\nu}\eta_{\rho\sigma}-\eta_{\mu\sigma}\eta_{\nu\rho})
+f^{ade}f^{bce}(\eta_{\mu\nu}\eta_{\rho\sigma}-\eta_{\mu\rho}\eta_{\nu\sigma})\Big].\nonumber\\
\label{eq:three-gluon-vertex-ordinary}
\end{align}
On the other hand, the second term of \eqref{eq:4-point}, arising only from VSR contributions, contains a huge number of terms.
Next, we define the total amplitude of the gluon's four-point function as
\begin{align}
\Omega^{abcd}_{\mu\nu\rho\sigma}\Big|_{\tiny\mbox{\textbf{Total}}}=
\sum\limits_{i=1}^{24}\Big\{\Omega^{abcd}_{\mu\nu\rho\sigma}\Big|_{\tiny\mbox{\textbf{(a)}}}^{(i)}
+\Omega^{abcd}_{\mu\nu\rho\sigma}\Big|_{\tiny\mbox{\textbf{(b)}}}^{(i)}
+\Omega^{abcd}_{\mu\nu\rho\sigma}\Big|_{\tiny\mbox{\textbf{(c)}}}^{(i)}
+\Omega^{abcd}_{\mu\nu\rho\sigma}\Big|_{\tiny\mbox{\textbf{(d)}}}^{(i)}
+\Omega^{abcd}_{\mu\nu\rho\sigma}\Big|_{\tiny\mbox{\textbf{(e)}}}^{(i)}
\Big\},
\end{align}
In order to have the leading VSR contributions, we keep only the VSR terms at the order of $(m^{2}/m_q^2)$ as
\begin{align}
\Omega^{abcd}_{\mu\nu\rho\sigma}\Big|_{\tiny\mbox{\textbf{Total}}}=
\Theta^{abcd}_{\mu\nu\rho\sigma}(p_1,p_2,p_3)\Big|_{\tiny\mbox{\textbf{(a)}}}
+m^{2}\Upsilon^{abcd}_{\mu\nu\rho\sigma}(p_1,p_2,p_3,n)\Big|_{\tiny\mbox{\textbf{Total}}}.
\label{eq:total-4-point}
\end{align}
At last, we remark that the ordinary contributions come only from the diagram (a), while the second part in \eqref{eq:total-4-point} includes all of the VSR terms containing leading contribution at $m^2$-order, originating from all of the five diagrams in Fig.~\ref{fig:fourpoint}.

Finally, in order to generate the complete induced ordinary Yang-Mills action, we should gather all of the ordinary Yang-Mills terms, which are at the order of $\frac{1}{|m_q|}$, in the amplitudes of $\langle GG\rangle$, $\langle GGG\rangle$ and $\langle GGGG\rangle$.
To this purpose, we consider Eqs.~\eqref{eq:pi}, \eqref{eq:3-gluon-vertex} and \eqref{eq:three-gluon-vertex-ordinary}.
Hence, by inserting them into \eqref{eq:1-4} and \eqref{eq:1-3}, we find the usual (non-VSR) Yang-Mills action as the following
\begin{align}
\Gamma^{\tiny\mbox{YM}}_{\tiny\mbox{eff}}[G]=\frac{icg^2}{\pi|m_q|}\int d^{3}x~\textrm{tr}(F_{\mu\nu}F^{\mu\nu}),
\end{align}
where $F_{\mu\nu}=\partial_{\mu}G_{\nu}-\partial_{\nu}G_{\mu}+ig[G_{\mu},G_{\nu}]$ and $c$ is a numerical factor.
This action is invariant under the infinitesimal non-abelian gauge transformations \eqref{eq105}, with $\kappa\rightarrow 0$.

\section{Modified field equation}
\label{sec6}

In this section, by completion, we intend to highlight the main features of the VSR effects in the ordinary non-abelian Chern-Simons theory, described by \eqref{CS-action}.
To this end, we consider only the VSR contributions at the leading order $(m^2/m^2_q)$ of the amplitudes $\langle GG\rangle$, $\langle GGG\rangle$, i.e. Eqs.~\eqref{eq:pi} and \eqref{eq:final-c}, respectively.
Thus, the relevant effective action at this order, for the positive value of $m_{q}$, is presented as below
\begin{align}
\Gamma_{\tiny\mbox{CS}}^{vsr} &=\frac{g^2}{8 \pi}\int d^{3}x ~\varepsilon_{\mu \nu \rho}
\Big( G^{a \mu} \partial^{\nu} G^{a \rho} + \frac{ig}{3}  f^{a b c} G^{a \mu}
G^{b \nu} G^{c \rho} \Big) \cr
 &+
  \frac{g^2}{32 \pi}
\Big(\frac{m^2}{m^2_q}\Big)\int d^{3}x ~\varepsilon_{\mu \nu \rho}n^{\rho}\Bigg[G^{a \mu}
\frac{1}{n.\partial}  (\Box G^{a \nu})
+
 (n.G^a) \frac{1}{(n.\partial)^2}
\Box \partial^{\nu} G^{a \mu} - G^{a \mu} \frac{1}{(n.\partial)^2} \Box \partial^{\nu} (n.G^a) \cr
&+ \frac{2ig f^{a
b c}}{3}
\Big(\frac{ \partial^{\nu}}{n.\partial} G^{a \mu} \Big) n.G^b \Big( \frac{\Box}{(n.\partial)^2} n.G^c \Big)\Bigg].
\end{align}
The field equations for the gauge field $G_{\mu}^{a}$ can be found from the variational principle $\delta\Gamma_{\tiny\mbox{CS}}^{vsr} =0$, resulting into
\begin{align}
&\varepsilon_{\mu \nu \rho}
\Big(2 \partial^{\nu} G^{a \rho}+ ig f^{a b c}
G^{b \nu} G^{c \rho}  \Big) \cr
 &+  \frac{1}{4}
\Big(\frac{m^2}{m^2_q}\Big)
\Bigg\{2n^{\rho}\Big[\varepsilon_{\mu \nu \rho}
\frac{1}{n.\partial}  (\Box G^{a \nu}) +
 \varepsilon_{\lambda \nu \rho}n_{\mu} \frac{1}{(n.\partial)^2}
\Box \partial^{\nu} G^{a \lambda}
-\varepsilon_{\mu \nu \rho}n^{\lambda} \frac{1}{(n.\partial)^2}
\Box \partial^{\nu} G^{a}_{\lambda}\Big] \cr
&+ \frac{2ig }{3}f^{abc}n^{\rho}
\Bigg[
\varepsilon_{\mu \nu \rho}n_{\beta}n_{\xi}\Big[\Big(\frac{ \partial^{\nu}}{n.\partial} G^{b\beta} \Big) \Big( \frac{\Box}{(n.\partial)^2} G^{c\xi} \Big)
+\Big(\frac{1}{n.\partial}  G^{b\beta}\Big) \Big( \frac{\partial^{\nu}\Box}{(n.\partial)^2} G^{c\xi} \Big)
+(\partial^{\nu}G^{b\beta}) \Big( \frac{\Box}{(n.\partial)^3} G^{c\xi} \Big) \cr
&+ G^{b\beta} \Big( \frac{ \partial^{\nu}\Box}{(n.\partial)^3} G^{c\xi} \Big)\Big]
-\varepsilon_{\lambda \nu \rho}n_{\mu}n_{\xi}
\Big(\frac{ \partial^{\nu}}{n.\partial} G^{b \lambda} \Big) \Big( \frac{\Box}{(n.\partial)^2} G^{c\xi} \Big) \cr
 &-
\varepsilon_{\lambda \nu \rho}n_{\mu}n_{\beta}\Big[
\Big(\frac{ \Box\partial^{\nu}}{(n.\partial)^3} G^{c\lambda} \Big) G^{b\beta}
+2\Big(\frac{\partial^{\alpha} \partial^{\nu}}{(n.\partial)^3} G^{c\lambda} \Big) \partial_{\alpha}G^{b\beta}
+\Big(\frac{ \partial^{\nu}}{(n.\partial)^3} G^{c\lambda} \Big)\Box G^{b\beta} \cr
&+ 2\Big(\frac{ \Box\partial^{\nu}}{(n.\partial)^2} G^{c\lambda} \Big) \Big(\frac{1}{n.\partial}G^{b\beta}\Big)
+4\Big(\frac{ \partial^{\alpha}\partial^{\nu}}{(n.\partial)^2} G^{c \lambda} \Big)\Big(\frac{\partial_{\alpha}}{n.\partial} G^{b\beta}\Big)
+ 2\Big(\frac{ \partial^{\nu}}{(n.\partial)^2} G^{c\lambda} \Big) \Big(\frac{\Box}{n.\partial}G^{b\beta}\Big) \cr
&+ \Big(\frac{ \Box\partial^{\nu}}{n.\partial} G^{c\lambda} \Big) \Big(\frac{1}{(n.\partial)^2}G^{b\beta}\Big)
+
2\Big(\frac{ \partial^{\alpha}\partial^{\nu}}{n.\partial} G^{c\lambda} \Big) \Big(\frac{\partial_{\alpha}}{(n.\partial)^2}G^{b\beta}\Big)
+
\Big(\frac{ \partial^{\nu}}{n.\partial} G^{c\lambda} \Big) \Big(\frac{\Box}{(n.\partial)^2}G^{b\beta}\Big)
\Big]\Bigg]
\Bigg\}=0.
\label{field_equation}
\end{align}
The first line corresponds to the ordinary non-abelian pure Chern-Simons equation of motion, and the other terms, which are proportional to $m^2$, indicate the VSR corrections.
For the source free case, it is well known that the ordinary Chern-Simons theory has $F_{\mu\nu}=0$, which leads to the pure gauge solutions (flat connections) as $G_{\mu}=\mathfrak{ g}^{-1} \partial_{\mu}\mathfrak{ g}$, where $\mathfrak{ g}$ is in the gauge group \cite{Dunne:1998qy}.
Naturally, this solution no longer holds for \eqref{field_equation} and should be complemented by VSR correction terms.
However, it is not an easy endeavour to obtain this VSR pure gauge solution since \eqref{field_equation} involves nonlocal and higher-derivative terms.


\section{Final remarks}
\label{conc}

In this paper, we have studied the one-loop effective action for a non-abelian gauge field, with the SU(3) gauge group, in (2+1)-dimensional spacetime with VSR group SIM(1). Our main interest was the analysis of the VSR modification in the non-abelian Chern-Simons terms.

In order to find the non-vanishing terms of the series $\Gamma[G]$, we started with the evaluation of the respective graphs contributing to the one-loop amplitude of $\langle GG\rangle$, corresponding to the free part of the gauge field effective action.
The one-loop result of $\langle GG\rangle$ was the same as the abelian case \cite{Bufalo:2020cst}, except for the overall color factor $Tr(T^aT^b)$.
 Actually, at the low-energy limit, the amplitude of $\langle GG\rangle$ contained the kinetic term of the Yang-Mills-Chern-Simons action and their VSR modifications. Besides, as we expected, in the non-VSR limit, $m\to 0$, the usual free parts of the Yang-Mills-Chern-Simons action were recovered.

In the next step, we have examined the relevant diagrams contributing to the one-loop amplitude of $\langle GGG\rangle$, corresponding to the cubic self-coupling part of the gauge field effective action.
This calculation in the low-energy limit led to the generation of the self-coupling terms in two different sectors: parity-even Yang-Mills piece and the parity-odd Chern-Simons part, as well as their VSR corrections.
We observed that Furry's theorem is not satisfied and indeed $\langle GGG\rangle$ led to the generation of cubic self-coupling terms in non-abelian case.

In order to complete the structure of the induced ordinary Yang-Mills action, it was necessary to generate the quartic part of the self-coupling terms. Therefore, we considered all of the respective one-loop diagrams related to the amplitude $\langle GGGG\rangle$. Our result demonstrated that the leading non-VSR contributions are exactly the quartic part of the self-coupling terms in the Yang-Mills sector at the low-energy limit.

Finally, in order to highlight the main features of the VSR effects in the non-abelian Chern-Simons action, we have considered the induced non-abelian Chern-Simons action with its leading VSR corrections.
 Next, we have obtained the equation of motion for the gauge field, containing the nonlocal and higher-derivative VSR terms.
 Taking into account the complexity of the equation of motion, we realized that the pure gauge solutions of the standard case are no longer satisfied in the presence of the VSR effects.

 \subsection*{Acknowledgements}
 R.B. acknowledges partial support from Conselho
Nacional de Desenvolvimento Cient\'ifico e Tecnol\'ogico (CNPq Projects No. 305427/2019-9 and No. 421886/2018-8) and Funda\c{c}\~ao de Amparo \`a Pesquisa do Estado de Minas Gerais (FAPEMIG Project No. APQ-01142-17).

\appendix

\section{Useful integral}
\label{apA}

In order to cope with the momentum integrals in VSR, we use the Mandelstan-Leibbrant prescription. A simple way to get the integrals was developed in \cite{Alfaro:2016pjw}. Here, we mention an important integral for our computations as below:
\begin{eqnarray}
\label{intprim}
I_{vsr}(a,b,q,\Delta)&=&\int d^d p \frac{1}{(p^2 + 2 p \cdot q - \Delta^2)^a} \frac{1}{(n \cdot p)^b}\nonumber\\
&=& (- 1)^{a + b} i \pi^{\omega} (- 2)^b \frac{\Gamma (a + b - \omega)}{\Gamma(a) \Gamma (b)} (\bar{n} \cdot q)^b I_{V}(a,b,q,\Delta),
\end{eqnarray}
with $\omega=d/2$ and
\begin{equation}
\label{iV1}
I_{V}(a,b,q,\Delta)=\int^1_0 d t t^{b - 1} \frac{1}{[\Delta^2 +q^2 - 2 (n \cdot q) (\bar{n} \cdot q) t]^{a + b - \omega}}.
\end{equation}
This is the basic integral. Other tensorial integrals can be obtained from \ref{intprim} by taking a derivative with respect to $p_\mu$.
\vspace{1cm}


\global\long\def\link#1#2{\href{http://eudml.org/#1}{#2}}
 \global\long\def\doi#1#2{\href{http://dx.doi.org/#1}{#2}}
 \global\long\def\arXiv#1#2{\href{http://arxiv.org/abs/#1}{arXiv:#1 [#2]}}
 \global\long\def\arXivOld#1{\href{http://arxiv.org/abs/#1}{arXiv:#1}}


\end{document}